\documentclass[twocolumn,amsmath,amssymb,prb,aps,showkeys,superscriptaddress]{revtex4-1}
\usepackage{graphicx}% Include figure files
\usepackage{dcolumn}% Align table columns on decimal point
\usepackage{bm}% bold math
\usepackage{amssymb}
\usepackage{graphicx}
\usepackage{amsmath}
\usepackage{xspace}
\begin{document}
\title{Phonon Fingerprints of CsPb$_2$Br$_5$ Single Crystals}

\author{V.~G.~Hadjiev}
\email{vhadjiev@uh.edu}
\affiliation{Texas Center for Superconductivity, University of Houston, Houston, Texas
77204, USA}
\affiliation{Department of Mechanical Engineering, University of Houston, Houston, Texas
77204, USA}
\author{C. Wang}
\affiliation{Department of Electrical and Computer Engineering, University of Houston, Houston, Texas 77204, United States}
\affiliation{Institute of Optoelectronic Information Material, School of Material Science and Engineering, Yunnan University, Kunming, Yunnan 650091, China}
\author{Y. Wang}
\affiliation{Department of Electrical and Computer Engineering, University of Houston, Houston, Texas 77204, United States}
\affiliation{Institute of Fundamental and Frontier Sciences, University of Electronic Science and Technology of China, Chengdu, Sichuan 610054, China}
\author{X. Su}
\affiliation{Department of Electrical and Computer Engineering, University of Houston, Houston, Texas 77204, United States}
\affiliation{School of Materials Science and Engineering, Chang\textsf{'}an University, Xi\textsf{'}an, Shaanxi 710061, China}
\author{H. A. Calderon}
\affiliation{Instituto Polit\'{e}cnico Nacional, ESFM, Departamento de Ciencia de Materiales, UPALM,  CDMX 07338, Mexico}
\author{F. Robles Hernandez}
\affiliation{Mechanical Engineering Technology, University of Houston, Houston, TX 77204, United States}
\author{Z. M. Wang}
\affiliation{Institute of Fundamental and Frontier Sciences, University of Electronic Science and Technology of China, Chengdu, Sichuan 610054, China}
\author{J. M. Bao}
\affiliation{Department of Electrical and Computer Engineering, University of Houston, Houston, Texas 77204, United States}
\affiliation{Institute of Fundamental and Frontier Sciences, University of Electronic Science and Technology of China, Chengdu, Sichuan 610054, China}
\affiliation{Materials Science \& Engineering, University of Houston, Houston, Texas 77204, United States}

\date{February 8, 2018}
% It is always \today, today,
% but any date may be explicitly specified

\begin{abstract}
  CsPb$_2$Br$_5$ is a stable, water-resistant, material derived from CsPbBr$_3$ perovskite and featuring two-dimensional Pb-Br framework separated by Cs layers.
  Both compounds can coexist at nanolength scale, which often produces
conflicting optical spectroscopy results.
  We present a complete set of polarized Raman spectra of nonluminescent CsPb$_2$Br$_5$ single crystals that
reveals the symmetry and frequency of nondegenerate Raman active phonons accessible from the basal (001) plane.
  The experimental results are in good agreement with density functional perturbation theory simulations,
which suggests that the calculated frequencies of yet unobserved double degenerate Raman
and infrared phonons are also reliable.
  Unlike CsPbBr$_3$, the lattice dynamics of CsPb$_2$Br$_5$ is stable as evidenced
by the calculated phonon dispersion.
  The sharp Raman lines and lack of a dynamic-disorder-induced central peak in the spectra at room temperature
indicate that the coupling of Cs anharmonic motion to Br atoms, known to cause the dynamic disorder in CsPbBr$_3$, is absent in CsPb$_2$Br$_5$.
\end{abstract}

%\pacs{81.05.Je 78.30.-j 63.20.D- 71.15.Mb }

\maketitle
%\keywords{}

%%%%%%%%%%%%%%%%%%%%%%%%%
\section{Introduction}
%%%%%%%%%%%%%%%%%%%%%%%%

   Cs-based lead halide perovskites have emerged as more temperature stable optoelectronic materials than
the hybrid organic-inorganic perovskite counterparts.\cite{Kulbak2015}
   The latter are notable for their
impressive power photoconversion efficiency of $> 20\%$,\cite{Yang2017a}
and potential applications as light emitting diodes\cite{Stranks2015} and thermoelectrics\cite{He2014}.
    Hybrid organic-inorganic and all-inorganic perovskites have comparable
photovoltaic performance.\cite{Kulbak2016}
    On the downside, both types of materials are water sensitive.
    In presence of water, CH$_3$NH$_3$PbI$_3$ degrades or forms hydrites\cite{Zhu2016} and CsPbBr$_3$ turns into CsPb$_2$Br$_5$\cite{Wang2018}.
    CsPb$_2$Br$_5$ is a water-resistant material\cite{Wang2018},\cite{Qiao2017} akin to the brightly photoluminescent (PL) CsPbBr$_3$\cite{Protesescu2015} but differing from the perovskites with its two-dimensional (2D)
Pb-Br framework separated by Cs layers.\cite{Powel1937}
    Both compounds are found or intentionally
prepared to coexists at nanolength scale.\cite{Palazon2017},\cite{Qiao2017}
    The corresponding nanocomposites\cite{Palazon2017} and CsPbBr$_3$/CsPb$_2$Br$_5$
 core/shell nanostructures\cite{Qiao2017} show stable PL and structural integrity.
    As to the PL properties of CsPb$_2$Br$_5$, reports are controversial: from emitting strong visible PL
 and even lasing capabilities\cite{Tang2017} to inherent PL inactivity\cite{Li2016},\cite{Zhang2018}.
    Visible PL was observed in nanocrystalline CsPb$_2$Br$_5$\cite{Yang2017} and in nanoplatelets\cite{Lv2018}.
    On the other hand, Refs. \onlinecite{Li2016} and \onlinecite{Dursan2017} report lack of PL
 in CsPb$_2$Br$_5$ nanocubes and single crystals, respectively.
    The PL controversy stems from the fact that CsPb$_2$Br$_5$ is an indirect
 band gap ($E_g\approx3$ eV) semiconductor\cite{Li2016},\cite{Zhang2018},\cite{Dursan2017} that is not supposed to emit PL in the range of 2.35--2.40 eV\cite{Tang2017},\cite{Yang2017},\cite{Lv2018}.
    A common trend in these experimental observations is that the forbidden PL in CsPb$_2$Br$_5$ is seen
 in nanostructures with complex morphology.
    The reasons for that could be remnant CsPbBr$_3$ embedded in CsPb$_2$Br$_5$, defects, crystal edge states or
 an interphase between the two materials.
    In most cases, attempts were made to resolve the controversy using x-ray diffraction (XRD)
 and differences in PL emissions of CsPb$_2$Br$_5$ and CsPbBr$_3$, but ambiguity remains.

    One of the pressing issues is to reconcile the results of DFT modeling, that is,
 the wide band gap and lack of reasons for emitting PL, with a particular crystal state of CsPb$_2$Br$_5$.
    The potential of Raman spectroscopy to resolve this problem has not been fully explored
 yet as only the Raman spectra of CsPbBr$_3$ are known\cite{Yaffe2017} but not those of CsPb$_2$Br$_5$.
    CsPbBr$_3$ undergoes two structural phase transitions with temperature: from cubic $Pm\bar3m$
 to tetragonal $P4/mbm$ at 403~K, and further to orthorhombic $Pbnm$ at 361~K\cite{Hirotsu1974}.
    Although at room temperature CsPbBr$_3$ is already in the lowest temperature phase, its
 Raman spectra show broad smeared phonon peaks and scattering background in
 a shape of a central peak (centered at zero cm$^{-1}$ Raman shift).\cite{Yaffe2017}
    The perovskite structure of CsPbBr$_3$ consists of apex-to-apex
 connected PbBr$_6$ octahedra in a 3D framework.
    A combined Raman and molecular dynamics (MD) simulation study\cite{Yaffe2017} of CsPbBr$_3$ show
 that the central peak is due to dynamic-disorder scattering from a head-to-head Cs anharmonic motion coupled to Br face expansion of PbBr$_6$ octahedra.
    The 2D Pb-Br framework in CsPb$_2$Br$_5$ is not connected along the $c$-axis but separated by Cs layers.
    Thus if the dynamic-disorder scattering mechanism proposed in Ref.~\onlinecite{Yaffe2017}
 is viable then we should not expect a central peak because Pb-Br layers in CsPb$_2$Br$_5$ lack bridging Br atoms.

    In this work, we present an original Raman study of
 CsPb$_2$Br$_5$ aimed to reveal the intrinsic vibrational properties of PL inactive single crystals.
    A complementary density-functional perturbation theory (DFPT) simulation was carried out
 for calculating the lattice dynamics in CsPb$_2$Br$_5$ and thereby to confirm
 the reliability of Raman experiment and structural purity of CsPb$_2$Br$_5$ crystals.
    We also predict the phonon frequencies of Raman active modes not seen yet experimentally
because of crystal morphology constraints.

%%%%%%%%%%%%%%%%%%%%
\section{Material preparation, characterization, and Raman experiment}
%%%%%%%%%%%%%%%%%%%%%%%%%%%

     CsPb$_2$Br$_5$ microplatelets were grown by conversion of CsPbBr$_3$ in pure water.\cite{Wang2018}
     CsPbBr$_3$ powders (micro-cubes) were first synthesized using a modified method
by mixing 0.5~M Pb(CH$_3$COO)$_2\cdot3$H$_2$O and 1~M CsBr in 48$\%$ HBr
solution at room temperature.\cite{Wang2018},\cite{Stoumpos2013}
     CsPb$_2$Br$_5$ was then synthesized by simply dropping CsPbBr$_3$ micro-cubes
in large quantity of water in a flask at room temperature.
     Orange CsPbBr$_3$ quickly turned
into white and precipitates at the bottom of the flask.
     The white precipitates, consisting of mainly platelet crystals, were taken out and dried for further study.
     XRD measurements revealed very pure phases of initial CsPbBr$_3$
and precipitated CsPb$_2$Br$_5$ materials.\cite{Wang2018},\cite{Stoumpos2013}

    The Raman scattering spectra of CsPb$_2$Br$_5$ were measured with a Horiba
JY T64000 triple spectrometer on samples
placed in an Oxford Instruments Microstat$^{He}$ optical cryostat.
    All spectra were recorded in backscattering configurations with incident and
scattered light propagating normal to the CsPb$_2$Br$_5$ crystal platelet surfaces.
    The backscattering configurations are given in Porto's notation: $A(BC)\bar A$, where $A$ and
$\bar A$ are the propagation directions of incident and scattered light, respectively,
and $B$ and $C$ are the corresponding light polarizations $\vec{e}_i$ and $\vec{e}_s$.

%%%%%%%%%%%%%%%%%%%%%%%%%%
\section{Experimental results}
%%%%%%%%%%%%%%%%%%%%%%%%%%

    CsPb$_2$Br$_5$ crystalizes in a body-centered tetragonal structure,\cite{Powel1937} space group
$I4/mcm$ (No. 140), with lattice parameters typically close to those originally reported in Ref.~\onlinecite{Cola1971}.
    The CsPb$_2$Br$_5$ crystals adopt a platelet morphology with large faces
parallel to the crystallographic (001) plane.\cite{Powel1937}
    The primitive unit cell (PC) contains two formula units of CsPb$_2$Br$_5$,
$N_{cell}=16$ atoms per PC with $3N_{cell}=48$ degrees of vibrational freedom.
    The irreducible representations of the $\Gamma$-point phonon modes
are $3A_{1g} + 2B_{1g}+3B_{2g}+ 5E_g + 2A_{1u}+ 5A_{2u} + 3B_{1u} + B_{2u} + 8E_u$,
and only the $A_{1g}$, $B_{1g}$, $B_{2g}$, and $E_g$ phonons are Raman active.\cite{Rousseau1981}
    The acoustic modes have $A_{2u}$ and $E_u$ symmetry, whereas $A_{1u}$, $B_{1u}$, and $B_{2u}$
are neither IR nor Raman active.
    The remaining 4$A_{2u}$ and 7$E_u$ modes can be observed in far-IR spectroscopy experiments.
    The Raman tensor, $\Re_S=|\alpha_{ij}|$ with $i,j=x, y, z$, of active modes $S=A_{1g}, B_{1g}, B_{2g}, E_g$, has the following non-zero components:
$\Re_{A_{1g}}(\alpha_{xx}=\alpha_{yy}=a, \alpha_{zz}=b)$, $\Re_{B_{1g}}(\alpha_{xx}=-\alpha_{yy}=c)$,
$\Re_{B_{2g}}(\alpha_{xy}=\alpha_{yx}=d)$,  $\Re_{E_{g,1}}(\alpha_{yz}=\alpha_{zy}=e)$, and $\Re_{E_{g,2}}(\alpha_{xz}=\alpha_{zx}=-e)$.\cite{Rousseau1981}
    The analysis of Raman scattering activity  $I_S = [\vec{e}_s \cdot \Re_S \cdot \vec{e}_i]^2$ suggests
that measurements in four back scattering configurations from the surface of a CsPb$_2$Br$_5$
platelet are enough to determine the symmetry of nondegenerate phonons.
    These are $Z(XY)\bar Z$ with $ I_{B_{2g}}\neq0$, $Z(X'X')\bar Z$
with $ I_{A_{1g}}\neq0$ and $I_{B_{2g}}\neq0$, $Z(X'Y')\bar Z$ with $ I_{B_{1g}}\neq0$,
and $Z(XX)\bar Z$ with $ I_{A_{1g}}\neq0$ and $ I_{B_{1g}}\neq0$, where $Z$ and $Z'$ are
parallel to [001] crystallographic direction, $X$ is along [100] and orthogonal to $Y$, $X'$ and
$Y'$ denote [110] and [1$\bar1$0] directions, respectively.

%%%%%%%%%%%%%%%%%%%%%%%%%%
\begin{figure} [htb]
\includegraphics[width=10.5cm]{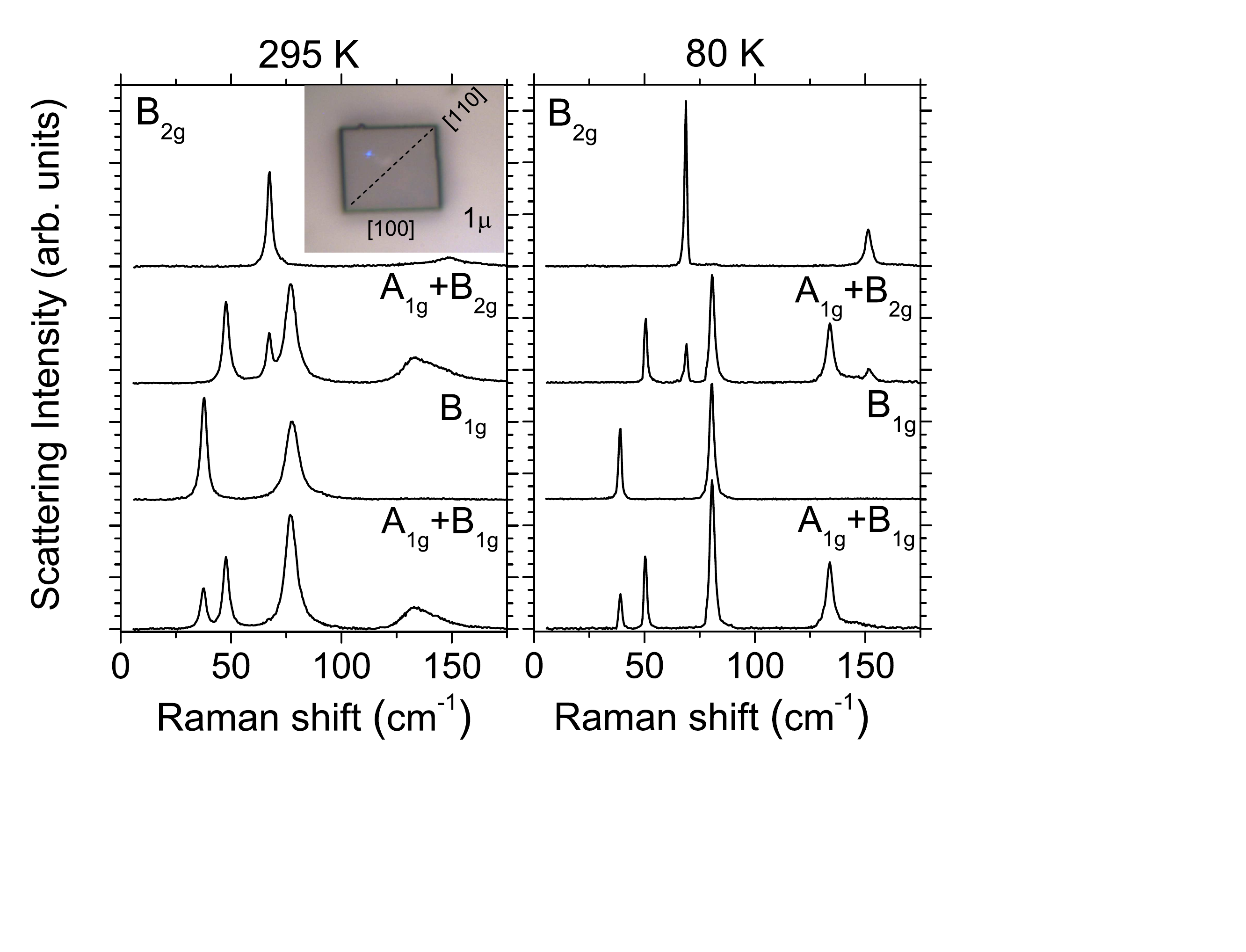}\\
\vspace{-20mm}
\caption{(Color online) Raman spectra of the CsPb$_2$Br$_5$ single crystal shown in the inset, excited with 632.8 nm laser line and measured in backscattering configurations with the laser
beam propagation direction along [001]. The crystal thickness is 0.5 $\rm{\mu}$m. The incident ($\vec{e}_i$) and scattered ($\vec{e}_s$) light polarization directions select $B_{2g}$ ($\vec{e}_i\ \| \ [100]$; $\vec{e}_s\ \| \ [010]$), $A_{1g}+B_{2g}$ ($\vec{e}_i\ \| \ [110]$; $\vec{e}_s\ \| \ [110]$), $B_{1g}$ ($\vec{e}_i\ \| \ [110]$; $\vec{e}_s\ \| \ [1\bar10]$), and $A_{1g}+B_{1g}$ ($\vec{e}_i\ \| \ [100]$; $\vec{e}_s\ \| \ [100]$).}
\end{figure}
%%%%%%%%%%%%%%%%%%%%%%%%%
    Figure~1 shows the polarized Raman spectra of the CsPb$_2$Br$_5$ crystal displayed in the inset,
measured in backscattering configurations from (001) crystal face.
    As seen in Fig.~1, the symmetry of $A_{1g}$, $B_{1g}$, and $B_{2g}$ phonons is
experimentally well established.
    The high single crystal quality of the sample is evidenced by the strongly polarized Raman spectra.
    The Raman spectra taken from a number of other CsPb$_2$Br$_5$ crystals including
those immersed in water support the phonon symmetry presented in Fig.~1.
    The frequency of all Raman phonons measured at 80~K are listed in Table~I.
    Notably, one $B_{2g}$ mode is missing in the strongly polarized Raman data.
    Having detected only two $B_{2g}$ phonons may confuse their symmetry assignment with that of the $B_{1g}$ modes.
    The results of DFPT calculations given in Table~I, however, are very helpful in this case and  confirm
the mode assignment in Fig.~1.
    As expected, none of the $E_g$ modes were observed in the measured crystal
due to the platelet crystal morphology.

    The measured CsPb$_2$Br$_5$ crystal show no PL emission in the visible light range
in accordance with the calculated electronic band
structure featuring a wide indirect band gap of $\approx$3 eV.\cite{Li2016},\cite{Dursan2017},\cite{Zhang2018}
    Thus we correlate the lack of PL emission to the single crystal nature of CsPb$_2$Br$_5$ sample.
    Similar conclusions are also done in Ref.~\onlinecite{Dursan2017}.

%%%%%%%%%%%%%%%%%%%%%%%%%%%
\begin{table}[htb]
    \caption{Experimental and DFPT Raman (R) and infrared (IR) phonon frequencies in CsPb$_2$Br$_5$ calculated using PAW and NC pseudopotentials. The corresponding lattice constants are $a=b=8.38~{\textrm \AA}$ and $c=15.27~{\textrm \AA}$ (PAW) and  $a=b=8.31~{\textrm \AA}$ and $c=15.26~{\textrm \AA}$ (NC). TO/LO splitting of $E_u$ modes is given for a phonon wavevector ${\bf q}\rightarrow0$ along $\Gamma$-M in the Brillouin zone.}
    \begin{ruledtabular}
        \begin{tabular}{|cccc|ccc|}
  mode      & exp.        & PAW           &NC            &  mode   & PAW & PAW \\
  sym.      & 80~K        & $\Gamma$-point & $\Gamma$-point & sym.   & $\Gamma$-point  & ${\bf q}\rightarrow0$ \\

            &             &                &     &  &       & along $\Gamma$-M\\
            &             &             &              &          &        & TO/LO \\
  R         &  cm$^{-1}$ &  cm$^{-1}$     &  cm$^{-1}$    &  IR     &  cm$^{-1}$      &  cm$^{-1}$ \\
  \hline
  $A_{1g}$  &  51         & 54          & 55                & $A_{2u}$  & 58  &   59   \\
  $A_{1g}$  &  81         & 82          & 83                & $A_{2u}$  & 73  &   73  \\
  $A_{1g}$  &  134        & 132         & 132               & $A_{2u}$  & 91  &   94   \\
  $B_{1g}$  &  39         & 41          & 42                & $A_{2u}$  & 141 &  153  \\
  $B_{1g}$  &  80         & 77          & 77                & $E_u$     & 18  &   18/38  \\
  $B_{2g}$  &  69         & 69          & 71                & $E_u$     & 46  &   46/53  \\
  $B_{2g}$  &             & 95          & 92                & $E_u$     & 60  &   60/64   \\
  $B_{2g}$  &  152        & 148         & 147               & $E_u$     & 73  &   73/84  \\
  $E_g$     &             &  36         & 31                & $E_u$     & 95  &   95/106   \\
  $E_g$     &             &  56         & 52                & $E_u$     & 112 &   112/116  \\
  $E_g$     &             &  70         & 64                & $E_u$     & 131 &   131/135  \\
  $E_g$     &             &  78         & 76                &           &     &          \\
  $E_g$     &             & 114         & 113               &           &     &        \\

  \hline
        \end{tabular}
    \end{ruledtabular}
\end{table}
%%%%%%%%%%%%%%%%%%%%%%%%%%%

%%%%%%%%%%%%%%%%%%%%%%%%%
\section{DFPT Calculation Details}
%%%%%%%%%%%%%%%%%%%%%%%%

%%%%%%%%%%%%%%%%%%%%%%%%%
\begin{figure}[htb]
\includegraphics[width=11cm]{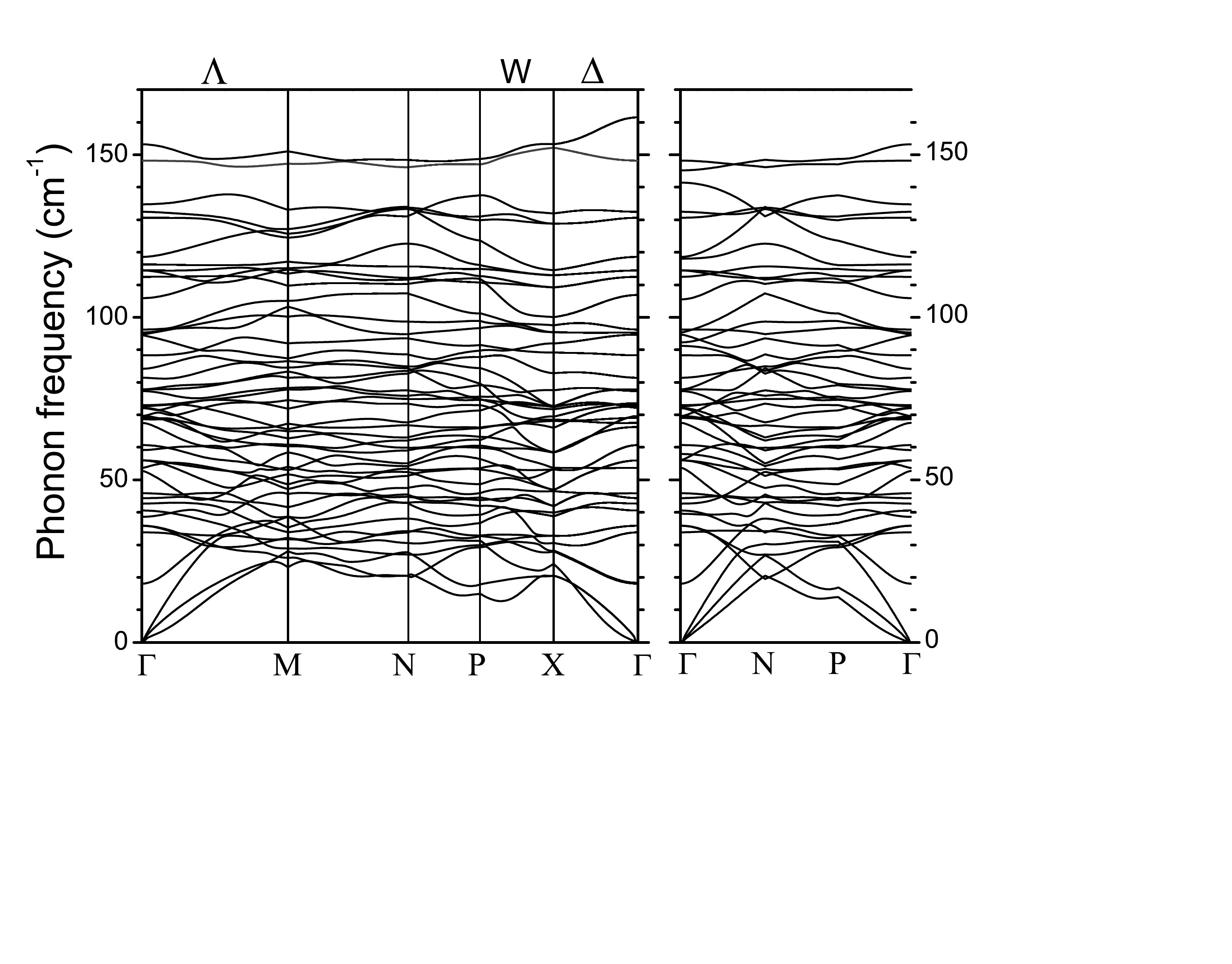}\\
\vspace{-25mm}
\caption{DFPT calculated phonon dispersion in CsPb$_2$Br$_5$ with TO/LO splitting included.}
\end{figure}
%%%%%%%%%%%%%%%%%%%%%%%%%
    The DFPT lattice dynamics calculations of CsPb$_2$Br$_5$ were performed within
the generalized-gradient approximation (GGA) with PBEsol functional \cite{Perdew2008}
using the DFPT code \cite{Baroni2001} as implemented in the Quantum Espresso (QE) suite \cite{Giannozzi2009}.
   In the calculations, we used the projector-augmented-wave (PAW) approach \cite{Blochl1994} with pseudopotentials
generated\cite{DalCorso2014} for use with QE.
    The geometry optimization of crystal structure, electronic band structure, and related properties
were calculated self-consistently (SCF) with 75 Ry kinetic energy
cutoff for the plane wave, 300 Ry charge density cut-off, SCF tolerance better than 10$^{-11}$, and 5.10$^{-6}$ Ry/au total residual force on atoms
over $4\times4\times4$ Monkhorst-Pack (MP) $k$-point grid.
    The dynamical matrices were calculated over $4\times4\times4$ MP $q$-point grid and used
after that for constructing the force constant matrix.
    Initial attempts to simulate the lattice dynamics of CsPb$_2$Br$_5$ at lower density $k$- and
$q$-point grids produced phonon dispersions with imaginary frequencies thus implying
possible inherent lattice instabilities as those seen in CsPbBr$_3$.
    However, increasing the density of both grids, although becoming quite computational demanding, resulted
in a stable lattice dynamics with calculated $\Gamma$-phonon frequencies
in a very good agreement with the low temperature experiment.
    The lattice constants calculated for the fully relaxed structure of CsPb$_2$Br$_5$
are $a=b=8.38~{\textrm \AA}$ and $c=15.27~{\textrm \AA}$.
    The DFPT calculations relax the size and shape of crystallographic unit cell
through minimization of all quantum mechanical forces in a static lattice, that is, at T=0~K.
    The calculated lattice constants are in a good agrement with the experimental ones
measured at room temperature: $a=b=8.48~{\textrm{\AA}}$ and $c=15.25~{\textrm \AA}$\cite{Cola1971}.
    We explored multiple combinations of functionals and pseudopotentials
in the calculations of lattice dynamics of CsPb$_2$Br$_5$ but none gave results as close to the experiment
as those produced by the GGA-PBEsol-PAW scheme.
    Only the calculations using the norm-conserved (NC) PBESol pseudopotentials gave reasonable values
for the lattice constants, $a=b=8.31~{\textrm \AA}$ and $c=15.26~{\textrm \AA}$, and $\Gamma$-point phonon frequencies (e.g. see Table I) but failed to produce a stable phonon dispersion.
    We purposely used another DFPT code\cite{Clark2005},\cite{Refson2006}
to calculate the non-resonant Raman intensity\cite{Porezag1996} using GGA-PBEsol-NC scheme as no such capability is available in QE.
%%%%%%%%%%%%%%%%%%%%%%%%%
\begin{figure}[htb]
\includegraphics[width=8.0cm, angle=00]{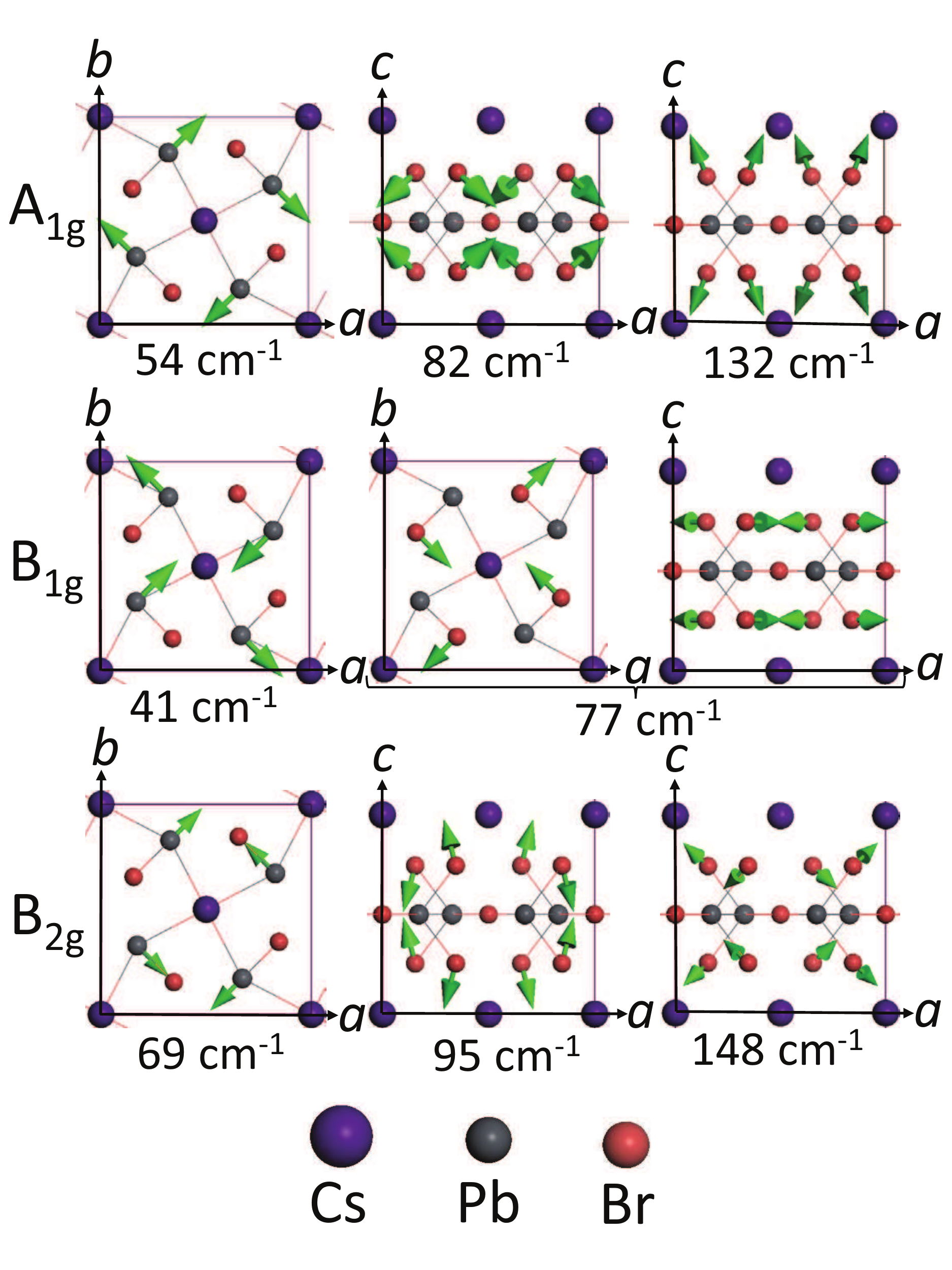}\\
\vspace{-5mm}
\caption{(Color online) Calculated atomic displacements for the non-degenerate Raman modes in CsPb$_2$Br$_5$. The axes $a$, $b$, and $c$ are along the crystallographic directions [100], [010], and [001], respectively.}
\end{figure}
%%%%%%%%%%%%%%%%%%%%%%%%%
%%%%%%%%%%%%%%%%%%%%%%%%%%%%%
\section{Discussion}
%%%%%%%%%%%%%%%%%%%%%%%%%%%%%%%

    Figure~2 displays the phonon dispersion of CsPb$_2$Br$_5$ calculated using GGA-PBESol-PAW.
    The TO/LO splitting of $E_u$ modes at $\Gamma$-point is also accounted for phonon
propagation towards the Brillouin zone boundaries points  M$(\frac{1}{2},\frac{1}{2},-\frac{1}{2})$,
X$(0,0,\frac{1}{2})$, N$(0,\frac{1}{2},0)$, and P$(\frac{1}{4},\frac{1}{4},\frac{1}{4})$ given in the primitive basis.
    The phonon dispersion in Fig.~2 displays phonon bands that are closely spaced in frequency and have
to be resolved by their eigenvectors.
    This is one of the reasons for making the lattice dynamics calculations
of CsPb$_2$Br$_5$ difficult and unstable.
    In addition, the out-of-plane acoustic modes in 2D materials have
quadratic dispersion and tend to produce negative frequencies around
$\Gamma$-point if the fast Fourier transformation
grid is not dense enough as shown in Ref.~\onlinecite{Iyikanat2017}
for CsPb$_2$Br$_5$ slab calculations.
    The quadratic dispersion effect in Fig.~2 is seen in the flattening of one of
the acoustic modes most pronounced at $\Gamma$-point along the $\Gamma$-P and $\Gamma$-X directions.

    As seen in Fig.~1 the predicted $B_{2g}$ mode at 95 cm$^{-1}$ lacks measurable Raman intensity.
    The calculations of Raman activity $I_S(\omega) = [\vec{e}_s \cdot (\partial \tilde{\alpha}/\partial Q_i) \cdot \vec{e}_i]^2$, where $\tilde{\alpha}$ is the polarizability tensor and $Q_i$ the normal mode coordinates, yielded $I_{B_{2g}}(95~{\textrm cm}^{-1})/I_{B_{2g}}(69~{\textrm cm}^{-1})=1.3\times10^{-3}$ and $I_{B_{2g}}(95~{\textrm cm}^{-1})/I_{B_{2g}}(148~{\textrm cm}^{-1})=8\times10^{-4}$ for the scattering configuration with $\vec{e}_i\ \| \ [100]$ and $\vec{e}_s\ \| \ [010]$.
    Therefore, the eigenvector of the 95~cm$^{-1}$ mode
produces vanishing modulation of the crystal polarizability.

    In Fig.~3, we show the calculated atomic displacements for all non-degenerate
Raman modes in CsPb$_2$Br$_5$.
    Comparing the experimental Raman intensities in Fig.~1
with the vibrational patterns in Fig.~3 we conclude that the intensity is strong for
all modes in which Br atoms move in-phase and predominantly in-plane in the Pb-Br layer.
    Apparently, this does not apply to the $B_{2g}$ mode at 95~cm$^{-1}$.

%%%%%%%%%%%%%%%%%%%%%%%%%
\begin{figure}[htb]
\includegraphics[width=9.5cm]{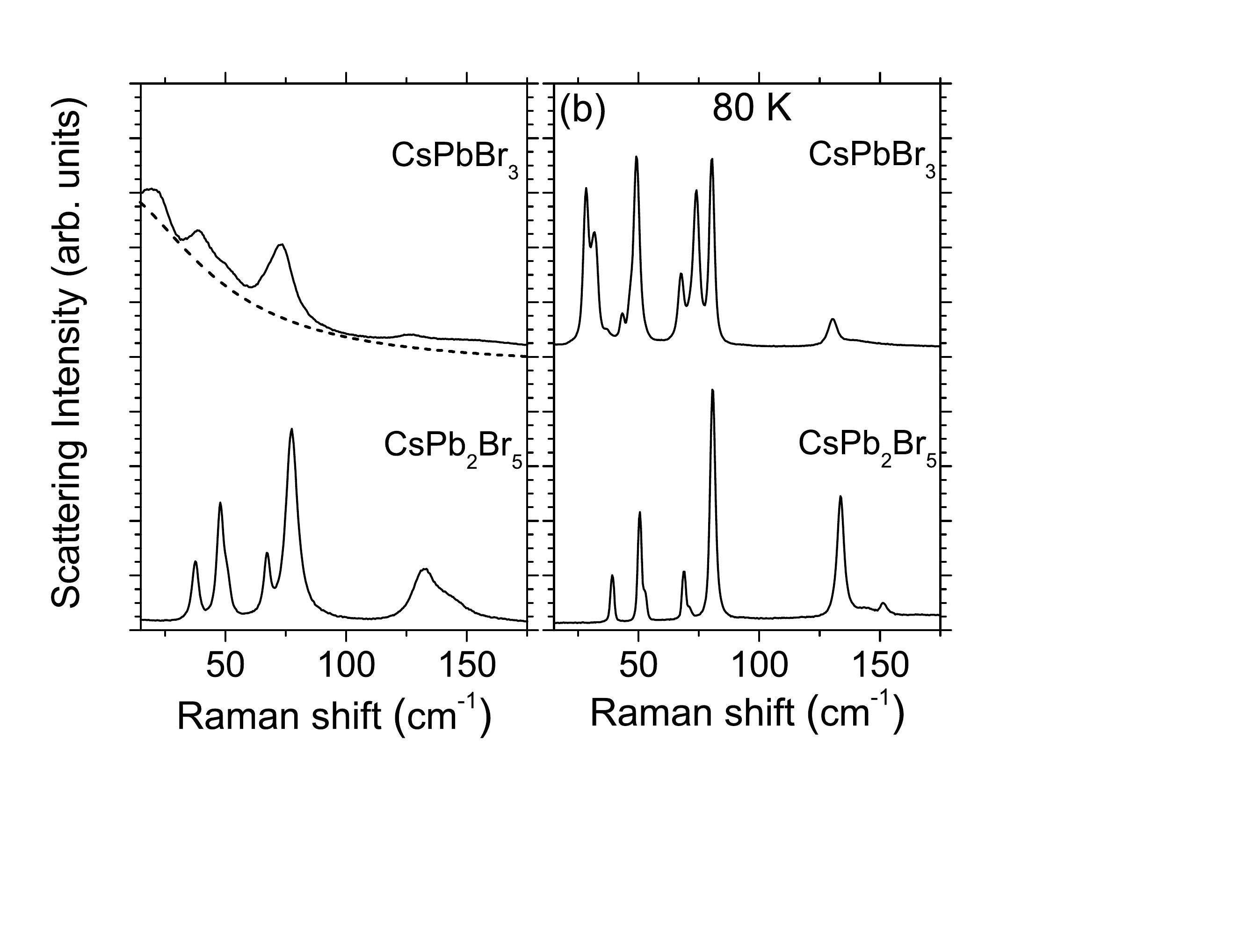}\\
\vspace{-15mm}
\caption{Comparison of non-polarised Raman spectra of CsPb$_2$Br$_5$ and CsPbBr$_3$. The dash curve in (a) is a guide to the eye that depicts the dynamic-disorder-induced central peak in CsPbBr$_3$\cite{Yaffe2017}.}
\end{figure}
%%%%%%%%%%%%%%%%%%%%%%%%%

    Figure~4 demonstrates the difference between the Raman spectra of CsPb$_2$Br$_5$ and CsPbBr$_3$
at room and low temperatures.
    The assignment of Raman phonons in  CsPbBr$_3$ is given in Ref.~\onlinecite{Yaffe2017}.
    We note, however, that the Raman spectrum of CsPbBr$_3$ shown in Fig.~4~(b) was measured at
a higher spectral resolution than that in Ref.~\onlinecite{Yaffe2017} and exhibits more spectral lines although not all of the expected by symmetry considerations.
    The presence of small amount of CsPbBr$_3$ as an impurity in CsPb$_2$Br$_5$ might
be challenging to detect at room temperature because of the smeared Raman features of CsPbBr$_3$.
    At low temperatures, however, the spectral range below 40 cm$^{-1}$ is free of CsPb$_2$Br$_5$ Raman lines
and the presence of CsPbBr$_3$ can be monitored through the Raman peaks at 28 cm$^{-1}$ and 32 cm$^{-1}$.
   We believe that Fig.~4 serves as a useful reference for the material characterization of both compounds
and will be helpful in detecting traces of CsPbBr$_3$ in CsPb$_2$Br$_5$.

%%%%%%%%%%%%%%%%%%%%%%%%%%%%%
\section{Conclusions}
%%%%%%%%%%%%%%%%%%%%%%%%%%%%%%%

    In summary, we have presented a comprehensive Raman scattering study of CsPb$_2$Br$_5$.
    The DFPT calculation results are in very good agreement with experimental phonon frequencies and symmetry.
    The latter gives us confidence that the calculated phonon dispersion and phonon related,
yet unmeasured, properties are also accessed properly.
    The present Raman study give an evidence that the lack of visible PL emission is
an intrinsic property of CsPb$_2$Br$_5$ single crystals.\\

%%%%%%%%%%%%%%%%%%%%%%%%%%%%%%%%%%%%%%%%
\begin{acknowledgments}

    VGH work was supported by the State of Texas through the Texas Center
for Superconductivity at the University of Houston (UH) and
in part with resources provided by the Center for Advanced Computing and Data Science (CACDS) at UH.
    JMB acknowledges support from the Robert A. Welch Foundation (E-1728).
    HAC acknowledges support from IPN (SIP, COFAA) for a sabbatical stay at the Molecular Foundry-LBNL. Work at the
Molecular Foundry was supported by the Office of Science, Office of Basic Energy Sciences,
of the U.S. Department of Energy under Contract No. DE-AC02-05CH1123.
\end{acknowledgments}

%%%%%%%%%%%%%%%%%%%%%%%%%%%%%%%%%%%%%%%%%%%

\end{document}